\PassOptionsToPackage{table}{xcolor}
\documentclass[11pt]{article}
\usepackage[letterpaper,left=0.85in,right=0.85in,top=0.90in,bottom=0.78in]{geometry}
\usepackage[T1]{fontenc}
\usepackage[utf8]{inputenc}
\usepackage{textcomp}
\usepackage{newtxtext,newtxmath}
\usepackage{amsmath}
\DeclareUnicodeCharacter{2248}{\ensuremath{\approx}}
\DeclareUnicodeCharacter{00D7}{\ensuremath{\times}}
\DeclareUnicodeCharacter{2264}{\ensuremath{\leq}}
\DeclareUnicodeCharacter{00B1}{\ensuremath{\pm}}
\DeclareUnicodeCharacter{2192}{\ensuremath{\rightarrow}}
\DeclareUnicodeCharacter{2212}{\ensuremath{-}}
\usepackage{xcolor}
\usepackage{graphicx}
\usepackage{longtable,booktabs,array,calc}
\usepackage{enumitem}
\usepackage{setspace}
\usepackage{titlesec}
\usepackage{fancyhdr}
\usepackage{microtype}
\usepackage{etoolbox}
\usepackage{hyperref}
\usepackage{xurl}
\hypersetup{hidelinks,hypertexnames=false,pdfcreator={pdfLaTeX},pdftitle={Noise Emissions from Hyperscale AI Data Centers: A Poorly Characterized Exposure at the Doorstep of Community Health Research}}

\titleformat{\section}{\normalfont\bfseries\fontsize{14}{17}\selectfont}{}{0pt}{}
\titlespacing*{\section}{0pt}{1.55em}{0.65em}
\titleformat{\subsection}{\normalfont\bfseries\fontsize{11.5}{14}\selectfont}{}{0pt}{}
\titlespacing*{\subsection}{0pt}{1.0em}{0.25em}

\newcommand{\papertitle}[1]{%
  \vspace*{0.34in}%
  \begin{center}%
    {\fontsize{18.5}{22.5}\selectfont #1\par}%
  \end{center}%
  \vspace{0.20in}%
}
\newcommand{\frontheadingcenter}[1]{{\bfseries\fontsize{11.5}{13.5}\selectfont #1\par}\vspace{0.15em}}
\newcommand{\frontheading}[1]{\par\vspace{0.55em}{\bfseries\fontsize{11.5}{13.5}\selectfont #1\par}\vspace{0.15em}}
\newenvironment{frontcenter}{\begin{center}\setstretch{1.06}\fontsize{10.4}{12.2}\selectfont}{\end{center}}
\newenvironment{abstractprose}{\begingroup\bfseries\setstretch{1.06}\fontsize{10.5}{12.2}\selectfont\noindent}{\par\endgroup}
\newenvironment{refblock}{%
  \begingroup
  \small
  \setstretch{1.12}%
  \setlength{\emergencystretch}{2.5em}%
  \begin{enumerate}[label=\arabic*.,leftmargin=2.15em,labelsep=0.55em,itemsep=0.62em,parsep=0pt,topsep=0.25em,partopsep=0pt,align=left]
}{%
  \end{enumerate}%
  \endgroup
}
\newcommand{\refauthors}[1]{\textbf{#1}}
\newcommand{\refurl}[1]{\par\nobreak\vspace{0.10em}{\footnotesize\urlstyle{tt}\url{#1}}}

\makeatletter
\def\maxwidth{\ifdim\Gin@nat@width>\linewidth\linewidth\else\Gin@nat@width\fi}
\def\maxheight{\ifdim\Gin@nat@height>\textheight\textheight\else\Gin@nat@height\fi}
\makeatother
\setkeys{Gin}{width=\maxwidth,height=\maxheight,keepaspectratio}

\begin{document}
\pagestyle{empty}
\papertitle{Noise Emissions from Hyperscale AI Data Centers: A Poorly
Characterized Exposure at the Doorstep of Community Health
Research}

\begin{frontcenter}
\frontheadingcenter{Author Information and Affiliations}

Davide Bray*, Francesca Dominici*

Department of Biostatistics, Harvard T.H. Chan School of Public Health,
655 Huntington Avenue, Building 2, 4th Floor, Boston, MA 02115, USA.

*Corresponding author

\end{frontcenter}

\frontheading{Abstract}

\begin{abstractprose}
The rapid build-out of hyperscale computing for artificial-intelligence
workloads has made data-center noise a recurring source of community
concern in the United States, yet calibrated, standards-grade acoustic
data remain scarce. This narrative review synthesizes direct
measurements, regulatory and legal records, investigative journalism,
and relevant engineering literature to assess what is currently known
and identify the measurements needed for credible health inference.
Available sources report heterogeneous acoustic values at residences,
property lines, and near-source positions, with individual readings from
39 to above 70 dB(A). Because location, metric, duration,
instrumentation, and provenance differ, these values are not directly
comparable community exposures, do not define a pooled distribution, and
do not support an aggregate comparison with nighttime limits. Among
cases with spectral detail, all but one reported continuous tonal noise
associated with complaints, most often low-frequency content near
70--140 Hz, that A-weighted metrics alone may under-represent.
Engineering evidence further suggests that cooling architecture may be a
major determinant of facility sound emission; the cited 15--30 dB span
is an equipment-level sound-power estimate, not a measured facility- or
residential-receptor difference, and remains a hypothesis pending
comparative field measurements. We propose a standards-based
field-measurement and modeling protocol designed to generate
frequency-resolved exposure data at named facilities. Such exposure
characterization is a necessary foundation for credible assessment of
community health effects and environmental justice implications.
\end{abstractprose}


\clearpage
\pagestyle{fancy}
\setcounter{page}{1}
\frontheading{Key Messages}

1. Public sources report heterogeneous acoustic values at residences,
property lines, and near-source positions, ranging from 39 dB(A) to
above 70 dB(A); differences in location, metric, duration,
instrumentation, and provenance preclude treating them as directly
comparable community exposures.

2. Complaints are dominated by continuous tonal noise (most often
low-frequency, ≈70--140 Hz) that A-weighted screening under-represents;
C-weighted levels, narrowband spectra, and tonality assessment are
necessary complements.

3. Cooling architecture is the strongest candidate predictor of facility
sound emission; the cited 15--30 dB span is an equipment-level
sound-power estimate, not a measured facility- or residential-receptor
difference.

4. A standards-grade measurement protocol is proposed as the
precondition for credible epidemiological studies of this rapidly
expanding exposure.

\section{AN EXPOSURE POORLY CHARACTERIZED IN THE HEALTH
LITERATURE}\label{an-exposure-poorly-characterized-in-the-health-literature}

Environmental-health research depends on being able to characterize
exposure before it can characterize risk. For the noise generated by
hyperscale artificial-intelligence data centers, that first step has not
been taken. Communities adjacent to new campuses in Tennessee,
Mississippi, Virginia, Arizona, and North Dakota have organized,
litigated, and rewritten local ordinances; where quantitative readings
exist at all, they come largely from consumer sound-level
meters.\textsuperscript{1--4} Local officials have set or contested
compliance thresholds without a calibrated, frequency-resolved picture
of what these facilities actually emit, or how that emission propagates
to nearby homes. The result is a public-health-relevant exposure that is
simultaneously the subject of active community concern and poorly
characterized in the peer-reviewed acoustic record; no peer-reviewed
data-center-specific epidemiological studies were identified in the
sources searched.

Three terms are used operationally throughout. A \emph{hyperscale data
center} is a campus-scale facility built and operated by, or for, a
major cloud or platform provider, typically with tens of megawatts or
more of installed IT power capacity; the newest AI campuses reach
hundreds of megawatts to more than a gigawatt. An \emph{AI data center}
is a facility whose dominant workload is machine-learning training or
inference on densely packed accelerator hardware, with rack power
densities roughly an order of magnitude above conventional enterprise
racks and sustained near-peak utilization; acoustically, this translates
into continuous, high-duty cooling and heat-rejection load rather than
the diurnally varying load of conventional facilities. \emph{Community
exposure} denotes sound levels attributable to facility operation at
residential receptors---outdoors at dwellings and, for low-frequency
content, indoors---as distinct from levels at property-line compliance
points or in the immediate vicinity of individual noise sources.

This is worth stating plainly, because it is easy to mistake volume of
reporting for quality of evidence. A structured search of the published
record (described below) turns up abundant journalism, legal filings,
and resident testimony, but essentially no peer-reviewed measurement
study---Tier 1, in the grading used here---of a named U.S. hyperscale
facility. The closest approximation is a consultant's
compliance-verification survey of Microsoft's SYD05 facility in New
South Wales, commissioned by the operator under a development-consent
condition: calibrated and standards-referenced, but a consulting report
(Tier 3) rather than peer-reviewed literature, and originating outside
the United States, which is itself a finding about where the incentives
in this space currently point.\textsuperscript{5} The purpose of this
review is to lay out what the existing evidence does and does not
support; to argue that cooling architecture and low-frequency tonal
content are the two variables around which any credible
exposure-assessment program should be organized; and to propose a
measurement and modeling protocol capable of producing the field's first
standards-grade dataset. Figure 1 maps the exposure pathway from
facility design to potential health outcomes, together with the evidence
currently available at each stage.

\begin{center}\includegraphics[width=0.86\textwidth]{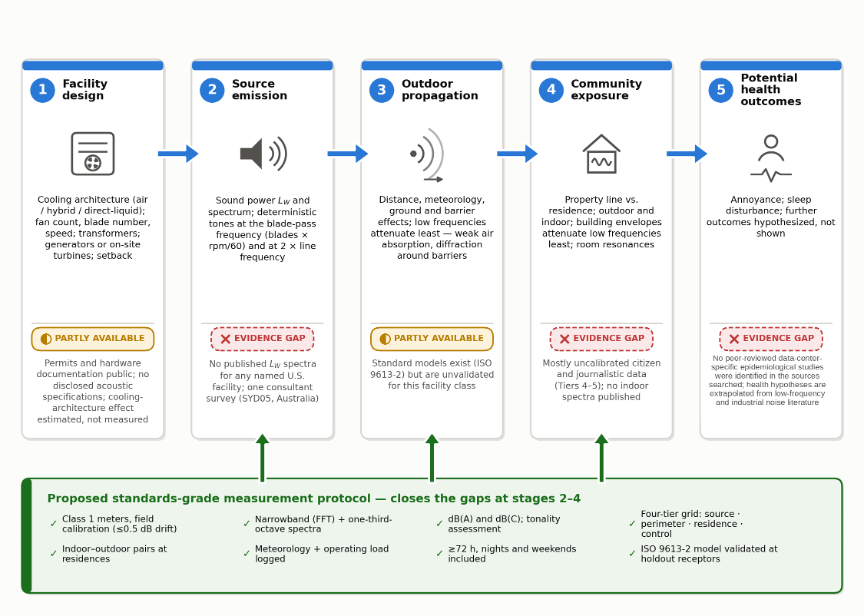}\end{center}

\textbf{Figure 1:} The exposure pathway from facility design to
potential health outcomes, with the current state of evidence at each
stage. Facility design parameters determine source sound power and its
spectrum, including deterministic tonal components at the fan blade-pass
frequency and at twice the power-line frequency. Outdoor propagation
favors low-frequency content, which building envelopes also attenuate
least, so indoor exposure at residences retains the low-frequency tonal
signature that dominates complaints. No stage yet rests on the
calibrated, frequency-resolved measurements that health inference
requires; the proposed protocol (bottom) is designed to supply them.

\section{EVIDENCE IDENTIFICATION AND
GRADING}\label{evidence-identification-and-grading}

This article is a narrative review, not a systematic review, and we make
its evidentiary basis explicit. We searched PubMed/MEDLINE, Scopus, and
Google Scholar through July 2026, combining terms for the source
category (data center, hyperscale, cloud, artificial intelligence) with
acoustic and health terms (noise, sound, acoustics, low-frequency,
tonal, hum, annoyance, sleep), without a starting date restriction.
Because almost none of the relevant material appears in indexed
journals, we additionally searched municipal and regulatory records
(including the Prince William County Data Center Ordinance Advisory
Group docket and the New South Wales planning portal), legal filings,
operator publications, and news archives, and followed citations forward
and backward from each source. We included every named hyperscale or
adjacent facility for which a quantitative acoustic level or spectral
signature has been publicly reported, and drew on measurement standards
and the low-frequency-noise health literature selectively, as context.
We did not perform formal screening, risk-of-bias assessment, or
quantitative synthesis of the kind a systematic review requires---no
such synthesis is possible on the current record, which is the gap this
review describes.

In place of a formal quality assessment, each reported value is graded
by evidentiary provenance on a five-tier scale, defined as follows. Tier
1 (T1): peer-reviewed measurement study using calibrated Class 1
instrumentation with published spectra. Tier 2 (T2): government,
regulatory, or legal record. Tier 3 (T3): professional
acoustic-consulting report using calibrated instrumentation, not
peer-reviewed. Tier 4 (T4): journalism reporting measurements. Tier 5
(T5): uncalibrated community or citizen data. The scale grades
provenance, not overall quality: a full multidimensional
assessment---instrument class, calibration status, documentation of
metrics and time response, background control, spectral
resolution---cannot be completed for most entries precisely because the
underlying documentation does not exist. No value summarized in this
review meets Tier 1.

\section{WHAT THE EXISTING RECORD ACTUALLY
SUPPORTS}\label{what-the-existing-record-actually-supports}

A note on acoustic quantities: sound power level
(\emph{L}\textsubscript{W}, dB re 1~pW) is a distance-independent
property of a source; sound pressure level (\emph{L}\textsubscript{p},
dB re 20~µPa) depends on where it is measured. A-weighted and C-weighted
sound pressure levels are written dB(A) and dB(C);
\emph{L}\textsubscript{Aeq} denotes the A-weighted equivalent continuous
level over a stated period, \emph{L}\textsubscript{AFmax} the maximum
A-weighted level with fast time weighting, \emph{L}\textsubscript{Cpeak}
the C-weighted peak level, and \emph{L}\textsubscript{WA} the A-weighted
sound power level. Near-source sound pressure levels at 1--2~m
characterize individual equipment items, not community exposure; they
cannot be compared directly with property-line or residential levels,
and none of these quantities is an estimate of another. Additional
descriptors used below are \emph{L}\textsubscript{Ceq} (the C-weighted
equivalent continuous level); \emph{L}\textsubscript{den} and
\emph{L}\textsubscript{night} (the day--evening--night and night-time
levels used in noise mapping and in the WHO guidelines); and
\emph{L}\textsubscript{eq}, \emph{L}\textsubscript{max}, and
\emph{L}\textsubscript{90} (the equivalent, maximum, and
90th-percentile-exceeded levels over a measurement period).

Grading the available measurements by provenance produces a striking
result: the best-documented American cases rest on residents' consumer
meters, mayoral walk-arounds, and consultant modeling, rarely on
calibrated Class 1 instrumentation with published
spectra.\textsuperscript{6,7} The widely cited figures for xAI's
Colossus complex in Memphis and Southaven, and for Amazon's data centers
adjacent to the Great Oak subdivision in Prince William County, come
from resident-operated meters of unknown weighting and time response,
used without the background characterization a standards-grade
assessment requires.\textsuperscript{1,2} They are indicative of a real
problem. They are not definitive exposure values, and a health study
that treats them as ground truth inherits their defects.

With that caveat attached, the documented cases provide a heterogeneous
set of reported acoustic values (Table 1). Individual values range from
39 dB(A) to above 70 dB(A), and some source records compare particular
readings with locally applicable or cited
limits.\textsuperscript{2,7,14} However, the values differ in location,
metric, duration, instrumentation, and provenance; they mix near-source,
property-line, and residential positions; attended and unattended
sessions; measured, modeled, and journalistic values; and calibrated and
uncalibrated instruments. They are not directly comparable community
exposures, do not support a pooled range or aggregate comparison with
nighttime limits, and we do not use them as such. More consistent than
the levels is the signature: complaints are driven disproportionately by
continuous tonal content rather than broadband A-weighted
exceedance---low-frequency in every case that reported spectral detail
except Chandler, where the documented tone lies at 630--1,000~Hz---and
formalized in Prince William County's move toward C-weighted and
octave-band criteria after residents documented persistent tones near
140~Hz (and, after partial mitigation, dominant peaks at 70--130~Hz with
residual components to ≈270~Hz) at Great Oak.\textsuperscript{7,12,13}
Near-source equipment measurements at operating hyperscale campuses do
exist but are rare in the public domain: the Microsoft SYD05 survey,
conducted with calibrated Brüel \& Kjær Type 2250/2270 meters referenced
against Australian and New South Wales guidance, remains the clearest
instance, reporting sound pressure levels for an enclosed emergency
diesel generator, under routine no-load testing, of 75--77~dB(A) at 1~m
falling to 52--58~dB(A) at 20--60~m, and transformer levels of 56~dB(A)
at 2~m.\textsuperscript{5}

What the record does not support is any confident cross-company
generalization. There is no direct-measurement corpus for Google, Meta,
or Oracle comparable even to the Amazon case, let alone to SYD05;
corporate sustainability and cooling-design materials from those firms,
and from Microsoft itself, are silent on measured, as-operated
acoustics, the SYD05 compliance report having been produced under a
regulatory consent condition rather than as voluntary
disclosure.\textsuperscript{5,20} A model built from public data alone
will therefore be strongest where Microsoft and Amazon evidence exists
and weakest elsewhere---an asymmetry that any subsequent health-exposure
work needs to carry forward explicitly as an uncertainty structure
rather than paper over.

\textbf{Table 1: Reported acoustic levels at named hyperscale and
adjacent data-center facilities, by evidentiary provenance}

\begingroup\footnotesize\setstretch{1.05}\setlength{\tabcolsep}{3pt}\renewcommand{\arraystretch}{1.15}\rowcolors{2}{gray!12}{white}\begin{longtable}[]{@{}
  >{\raggedright\arraybackslash}p{(\columnwidth - 6\tabcolsep) * \real{0.1444}}
  >{\raggedright\arraybackslash}p{(\columnwidth - 6\tabcolsep) * \real{0.1675}}
  >{\raggedright\arraybackslash}p{(\columnwidth - 6\tabcolsep) * \real{0.5583}}
  >{\raggedright\arraybackslash}p{(\columnwidth - 6\tabcolsep) * \real{0.1299}}@{}}
\toprule\noalign{}
\begin{minipage}[b]{\linewidth}\raggedright
\textbf{Facility}
\end{minipage} & \begin{minipage}[b]{\linewidth}\raggedright
\textbf{Location}
\end{minipage} & \begin{minipage}[b]{\linewidth}\raggedright
\textbf{Reported level / signature}
\end{minipage} & \begin{minipage}[b]{\linewidth}\raggedright
\textbf{Tier}
\end{minipage} \\
\midrule\noalign{}
\endhead
\bottomrule\noalign{}
\endlastfoot
xAI Colossus & Memphis, TN / Southaven, MS & Federal class action
alleges \textgreater70~dB(A) at residential property lines near the
Southaven turbine plant;\textsuperscript{8} resident meter readings
reaching the 70s outdoors;\textsuperscript{1} mayoral readings
39--70~dB(A);\textsuperscript{6} continuous mixed signature of
high-pitched whine with low-frequency rumble and tonal
hum;\textsuperscript{1,9} post-mitigation compliance reported by the
mayor, disputed by residents (news spot checks
42--65~dB(A))\textsuperscript{6,10} & T2/T4/T5 \\
Amazon Great Oak & Prince William Co., VA & Up to 65~dB(A) at
night;\textsuperscript{2} 69.4~dB(A) evening reading
(2022);\textsuperscript{11} 70--140~Hz tonal
components;\textsuperscript{7,12,13} ≈10~dB(A) reduction after fan
retrofit (≈US \$30 million, community estimate), from ≈58--65~dB(A) to
≈48--55~dB(A);\textsuperscript{2,13} post-mitigation C-weighted averages
≈60--63~dB(C)\textsuperscript{12} & T2/T4/T5 \\
CyrusOne Phoenix & Chandler, AZ & Ambient near data-center sites
≈45--50~dB per a city official;\textsuperscript{14} persistent tonal hum
documented since 2018, measured as a pure tone at
630--1,000~Hz\textsuperscript{3,15} & T4 \\
Vantage VA2 & Sterling (Loudoun Co.), VA & Readings of 51--53~dB(A)
observed by a county supervisor at a residential property line
(ordinance limit 55~dB(A));\textsuperscript{16,17} campus powered by
continuously running on-site gas turbines;\textsuperscript{17,18}
residents report tonal hum and low-frequency
vibration\textsuperscript{17} & T4 \\
Atlas Power & Near Williston, ND & Resident sound-level readings
disputed by the operator as uncalibrated;\textsuperscript{4} dominant
118~Hz low-frequency tone identified by a consultancy's remote
narrowband analysis of uncalibrated broadcast audio, consistent with
six-blade fans at ≈1,180~rpm\textsuperscript{19} & T4/T5 \\
Microsoft SYD05 & Kemps Creek, NSW, AU & Enclosed emergency generators
(no-load testing) 75--77~dB(A) at 1~m; AHUs 59~dB(A) at the external
louver; transformers 56~dB(A) at 2~m\textsuperscript{5} & T3 \\
\end{longtable}\endgroup

T1: peer-reviewed measurement study using calibrated Class 1
instrumentation with published spectra (no entry qualifies); T2:
government, regulatory, or legal record; T3: professional
acoustic-consulting report (calibrated instrumentation, not
peer-reviewed); T4: journalism reporting measurements; T5: uncalibrated
community or citizen data. A single case may carry multiple tiers;
grading follows the provenance of the underlying data (the Atlas Power
entry is T4/T5 because it rests on consultant analysis of uncalibrated
broadcast audio).

All values are sound pressure levels at the stated or implied positions,
as reported by the cited sources. Locations, metrics, durations,
instrumentation, and provenance are heterogeneous---near-source,
property-line, and residential; attended and unattended; calibrated and
uncalibrated---so the values are not directly comparable community
exposures and must not be read as a pooled distribution or aggregate
comparison with limits. Superscript numerals link each value to its
source. The table lists the cases located by the search described in the
text and is not claimed to be exhaustive. AHU: air-handling unit; rpm:
revolutions per minute.

\section{COOLING ARCHITECTURE AS A CANDIDATE KEYSTONE
VARIABLE}\label{cooling-architecture-as-a-candidate-keystone-variable}

If a single design specification had to predict a facility's off-site
noise, engineering considerations point to cooling architecture. The
reason is physical rather than statistical: equipment-level sound-power
estimates for the cooling technologies considered here span
approximately 15--30~dB; this is not a measured contrast between
operating facilities or at residential receptors. Air-cooled chillers
and evaporative cooling towers move enormous volumes of air through
fans, producing sound power levels (\emph{L}\textsubscript{W}) on the
order of 95--120~dB re 1~pW, with the energy of the dominant
propeller-fan sources concentrated in the 63--250~Hz octave
bands---exactly the low frequencies that propagate farthest and resist
mitigation.\textsuperscript{21--23} Immersion cooling eliminates server
fans, and direct-to-chip designs reduce or, in fanless implementations,
eliminate them; both sharply reduce the air volume that must be moved
per unit of heat rejected, although fan-driven outdoor heat rejection
generally remains at reduced scale. Vendor and trade-press accounts
report in-hall sound pressure levels falling from roughly 85--95~dB(A)
in air-cooled aisles to 65--75~dB(A) in liquid-cooled
facilities---occupational figures that bear on the server room rather
than the property line, and that are not comparable with the sound-power
values above.\textsuperscript{24,25} For the dominant propeller-fan
cooling-tower case, sound power scales approximately as
\emph{L}\textsubscript{W} ≈ 95 + 10 log\textsubscript{10}(fan hp)~dB,
with the A-weighted sound power level roughly 9~dB
lower.\textsuperscript{22,26} The 15--30~dB span cited in the Abstract
is a bounding equipment-level sound-power estimate of the difference
between the noisiest air-cooled configurations and designs that
eliminate or drastically shrink the major fan arrays; it is assembled
from equipment-level sound-power estimates, not from comparative
measurements across operating facilities differing only in cooling
design. It should not be interpreted as a measured facility contrast or
a residential-receptor difference. Direct facility-level comparisons do
not yet exist; producing them is one purpose of the protocol proposed
below.

The implication for exposure-assessment design follows directly: cooling
architecture is the natural primary stratification variable for any
comparative health-exposure study, with air-cooled, hybrid, and
direct-liquid-cooled campuses compared at matched IT load, and with
continuously running gas-turbine sites---an emerging and distinct
category exemplified by the turbine-powered Vantage campus in Loudoun
County, Virginia, and by the turbine plant serving xAI's Memphis
campuses from Southaven, Mississippi---treated separately, since they
carry an elevated, continuous low-frequency load more characteristic of
a power plant than a server farm.\textsuperscript{9,17,18}

\section{WHY THE STANDARD METRIC, ALONE, MISSES THE EXPOSURE THAT
MATTERS}\label{why-the-standard-metric-alone-misses-the-exposure-that-matters}

The defining acoustic problem at these facilities is not peak loudness
but continuous, low-frequency tonal content, and this is the point at
which acoustics and health assessment become inseparable. Two
deterministic source families dominate. Transformers emit tonal energy
concentrated near twice the local power-line frequency (120~Hz on the
60~Hz North American grid, 100~Hz in 50~Hz systems) with harmonics at
even multiples of line frequency (240 and 360~Hz in North America),
produced chiefly by core magnetostriction with additional load-dependent
components from winding electromagnetic forces.\textsuperscript{27} The
hum is therefore neither a strictly pure tone nor fully
load-independent, but its frequency signature is stable and narrow;
levels for large units are typically on the order of 60--75~dB(A) at the
0.3--2~m contours used in transformer sound-level test codes, and higher
for the largest units, often nominally compliant on A-weighted criteria
yet attracting a tonal adjustment of 3--6~dB under standard
assessment.\textsuperscript{27,28} Fans generate a blade-pass tone at
\emph{f}\textsubscript{BPF} = (blades × rpm)/60; a six-blade fan at
1,180~rpm produces a 118~Hz tone. When acoustics consultants isolated a
dominant 118~Hz tone by narrowband analysis of broadcast audio from a
North Dakota facility, they could infer from the tone, combined with
standard direct-drive motor speeds, a six-blade fan at approximately
1,180~rpm---an illustration that these tonal frequencies are
deterministic functions of engineering
specifications.\textsuperscript{19,29}

Low-frequency energy behaves badly for every standard assessment axis
simultaneously: it propagates farthest, since atmospheric absorption is
weak at low frequency; it diffracts around barriers and berms, whose
attenuation is greatest at mid and high frequencies and far smaller at
low frequencies; and it is de-emphasized by A-weighting itself, so a
facility can be nominally compliant on dB(A) while producing exactly the
penetrating hum residents describe.\textsuperscript{30} A dB(C) − dB(A)
difference approaching or exceeding 20~dB, the formal preliminary
screening step of DIN 45680, signals that A-weighted levels alone cannot
adequately describe the exposure at that receptor and must be
supplemented by C- or Z-weighted levels, narrowband spectra, and
tonality assessment.\textsuperscript{30} Tonal audibility should be
assessed by the objective narrowband engineering method of ISO 1996-2
(Annex J), which feeds tonal adjustments of 3--6~dB into rating levels
under ISO 1996-1; the one-third-octave survey method (Annex K) is a
screening tool whose audibility criteria are far less sensitive at low
frequencies (a 15~dB band-level difference is required in the bands up
to 125~Hz, versus 8~dB at 160--400~Hz and 5~dB at 500~Hz and above), so
narrowband analysis is required precisely in the band that matters
here.\textsuperscript{28,31}

None of this makes dB(A) the wrong metric. A-weighted indicators remain
indispensable: virtually the entire exposure--response evidence base in
environmental-noise epidemiology, including the relations underlying the
WHO Environmental Noise Guidelines, is expressed in A-weighted terms
such as \emph{L}\textsubscript{den} and \emph{L}\textsubscript{night},
and abandoning them would disconnect any new work from that accumulated
evidence.\textsuperscript{32} The argument here is narrower: for sources
whose emission is dominated by continuous low-frequency tonal content,
dB(A) alone is insufficient, and exposure assessment should report C- or
Z-weighted levels, full spectra, and tonality alongside it. For an
environmental-health readership, the consequence is direct: dB(A)
compliance, the metric underlying most current municipal ordinances, is
not on its own a reliable proxy for the exposure residents are actually
reporting, and a health-outcome study that uses dB(A)-only exposure
classification for these facilities risks systematic exposure
misclassification concentrated exactly where the annoyance and complaint
signal is strongest.

\section{TOWARD A STANDARDS-GRADE EXPOSURE
SURFACE}\label{toward-a-standards-grade-exposure-surface}

Closing this gap does not require new acoustic theory; it requires
applying existing, well-established standards to facilities that have so
far mostly been assessed with consumer equipment. We propose a
measurement and modeling program built on two components.

\emph{Field protocol.} A defensible campaign requires a four-tier
measurement grid: near-source attended points at dominant façades,
generators, and transformers; unattended perimeter stations at the most
exposed property-line sectors; residential receptor stations at the
nearest homes; and, critically, a background/control station, since even
the SYD05 survey could not characterize the nearest residential
receivers by plant-only measurement because traffic and other industrial
noise dominated at long range.\textsuperscript{5} Each element of the
protocol responds to a specific, documented failure mode of the existing
evidence base:

1. \emph{Instrumentation and calibration.} Class 1 integrating meters
with one-third-octave capability, logging \emph{L}\textsubscript{Aeq},
\emph{L}\textsubscript{Ceq}, \emph{L}\textsubscript{AFmax},
\emph{L}\textsubscript{Cpeak}, and full band spectra---because the Great
Oak record shows that dB(A) alone misses the lived experience of
continuous low-frequency hum; and field-calibration checks before and
after every attended session and at deployment and retrieval of every
unattended station, with a drift acceptance criterion of ≤0.5~dB---the
practice followed in the SYD05 survey but absent from essentially all
community data collected to date.\textsuperscript{5,12}

2. \emph{Narrowband analysis.} FFT spectra with resolution of 1~Hz or
finer at attended and receptor positions: the one-third-octave survey
criteria deliberately demand far larger band-level differences in the
low-frequency bands (15~dB up to 125~Hz versus 5--8~dB above), so weak
but clearly audible tones in the ≈70--140~Hz band at issue are found
reliably only by the narrowband method.\textsuperscript{31}

3. \emph{Meteorology and operating conditions.} Concurrent logging of
wind speed and direction, temperature, humidity, and cloud cover, since
outdoor propagation is strongly refraction-dependent and ISO 9613-2
predictions assume moderately favorable conditions;\textsuperscript{33}
and documentation of facility operating state---utility power draw,
cooling stage, generator-test schedules---with operator cooperation
where possible, since emission from load-dependent sources cannot
otherwise be interpreted.

4. \emph{Event exclusion.} Pre-specified criteria excluding transients
(aircraft, road-traffic pass-bys, animal and human activity),
precipitation, and intervals with wind above 5~m/s at the microphone,
with excluded intervals documented rather than silently removed---an
exclusion protocol whose absence is one of the defects that makes
existing community data unusable for exposure assessment.

5. \emph{Duration.} Unattended stations running at least 72 hours and
ideally five to seven days including a weekend, because complaint
reports consistently emphasize night and weekend conditions.

6. \emph{Indoor--outdoor pairs.} Paired outdoor-façade and
indoor-bedroom measurements at consenting residences, including corner
and mid-room positions for low-frequency assessment, because building
envelopes attenuate low frequencies least, indoor room modes can amplify
them, and sleep-relevant exposure occurs indoors.\textsuperscript{30,34}

7. \emph{Background characterization and model validation.} The
dedicated control station, supplemented where operator cooperation
allows by plant-quiescent intervals; and holdout receptor positions at
which ISO 9613-2 predictions are compared with measurement, with
residuals reported by frequency band.\textsuperscript{33}

\emph{Modeling.} A source-inventory-plus-propagation model, using ISO
9613-2 for outdoor propagation and back-calculating source sound power
from near-source data where only sound pressure is publicly available,
provides the most transparent regulator-facing method. ISO 9613-2 states
an accuracy of about ±3~dB within its validity conditions; allowing
additionally for source-inventory and meteorological uncertainty, we
adopt realistic field uncertainties of roughly ±3--5~dB at
well-characterized source-adjacent points and ±5--8~dB at
receptors.\textsuperscript{33} A semi-empirical fan-and-louver submodel,
coupling a thermal balance (IT heat load → required airflow → fan speed)
to AMCA fan-sound relationships, is the highest-payoff addition for
campuses with large exhaust arrays, since it also predicts blade-pass
tonal frequency directly from public fan
specifications.\textsuperscript{29}

A first study built on this footing should establish at least a week of
continuous night data at three to five named facilities spanning cooling
types, reporting \emph{L}\textsubscript{eq}, \emph{L}\textsubscript{90},
\emph{L}\textsubscript{max}, \emph{L}\textsubscript{night}, and full
spectra in both dB(A) and dB(C), paired with each site's disclosed
cooling type, fan inventory, transformer count, and setback. That
dataset would allow the \emph{L}\textsubscript{WA} ≈ 86 + 10
log\textsubscript{10}(fan hp)~dB scaling relation to be tested for the
first time against measured levels at facilities of this specific class,
rather than assumed from decades-old cooling-tower
literature.\textsuperscript{22,23}

\section{WHY THIS IS AN ENVIRONMENTAL-HEALTH PROBLEM, NOT ONLY AN
ACOUSTICS
PROBLEM}\label{why-this-is-an-environmental-health-problem-not-only-an-acoustics-problem}

The case for treating this as environmental-health infrastructure,
rather than a purely engineering question, rests on three observations.
First, the exposure is plausibly biologically relevant on the same
low-frequency and tonal-annoyance pathways established in the
industrial-noise literature, where a mature body of work has linked
low-frequency and tonal exposure to annoyance, and, less robustly, to
self-reported sleep disturbance, at levels not captured by overall dB(A)
ratings.\textsuperscript{34,35} Second, the siting pattern of these
facilities, driven by land cost, power availability, and tax incentives,
has in documented cases, most prominently the turbine plant and campuses
serving xAI in South Memphis and Southaven, placed them adjacent to
lower-income and historically industrial
neighborhoods,\textsuperscript{1,9} raising the same
environmental-justice questions that have structured decades of work on
other locally undesirable land uses; the exposure surface needed to
evaluate that distributional question does not yet exist. Third, the
pace of construction means the exposure window is opening now:
community-level exposure to this specific noise signature is a matter of
which facilities are already operating, with new campuses reaching full
power within a construction cycle of one to two years.

The health-evidence base for this exposure must be stated in three
layers. For environmental noise generally, the WHO Environmental Noise
Guidelines and the systematic reviews underlying them establish
exposure--response relationships for annoyance and sleep disturbance,
derived principally from road, rail, and aircraft noise, with
cardiovascular and metabolic effects likewise best established for
transportation sources.\textsuperscript{32,36--38} For low-frequency
noise specifically, a systematic review of observational studies found
consistent associations with annoyance and suggestive but
methodologically limited evidence for sleep-related and other
outcomes,\textsuperscript{39} and laboratory and field studies indicate
that low-frequency and tonal content increases annoyance and can affect
sleep quality at levels below those implied by A-weighted
criteria.\textsuperscript{40,41} For data-center noise specifically, no
peer-reviewed data-center-specific epidemiological studies were
identified in the sources searched. In particular, we identified no
cohort, cross-sectional, or panel study examining health outcomes in
relation to measured community exposure to data-center noise. The health
relevance asserted in this review is therefore an
extrapolation---defensible because the exposure signature (continuous,
tonal, low-frequency, and nocturnal) matches the features the
low-frequency-noise literature identifies as most problematic, but an
extrapolation nonetheless, and we present it as hypothesis rather than
established risk.

When health studies do become possible, exposure characterization will
not be the only design problem. Facility siting correlates with
socioeconomic position, housing stock, and pre-existing environmental
burden, so candidate confounders and effect modifiers include
socioeconomic status and housing quality (insulation, glazing, bedroom
orientation); co-exposures generated by the facilities themselves,
including diesel-generator testing, gas-turbine combustion emissions,
construction activity, traffic, and light at night; pre-existing ambient
noise from highways and industry; individual noise sensitivity; and
attitudes toward the operator---perceived economic benefit, procedural
fairness in siting---which are well-documented non-acoustic moderators
of noise annoyance.\textsuperscript{36} Residential self-selection and
differential mobility (who can afford to move away) operate in the same
direction as the environmental-justice concern itself. A credible design
will need to measure these alongside exposure rather than adjust for
them as an afterthought.

This creates an unusual opportunity and an unusual obligation. It is
rare in environmental-health research to be able to characterize an
emerging exposure before, rather than after, a large body of
health-outcome literature has been built on inadequate exposure data.
Speculative frameworks linking low-frequency data-center noise to health
outcomes are already circulating outside the peer-reviewed
literature---precisely the pattern, outcome speculation preceding
exposure characterization, that a field should try to avoid. The
sequencing matters: exposure has to be characterized to instrument-grade
standards first, because health inference is only as good as the
exposure it rests on.

\section{LIMITATIONS}\label{limitations}

This review has limitations that follow directly from its subject.
First, it is a narrative review; although the search strategy is
described above, we did not apply systematic-review screening or formal
quality assessment, and the selection of gray-literature sources
necessarily involved judgment. Second, the quantitative values
summarized here are heterogeneous in location, metric, duration,
instrumentation, and provenance; individual reported values span 39 to
above 70~dB(A), but they do not support a pooled range, a directly
comparable community-exposure distribution, or an aggregate comparison
with nighttime limits. Third, the record is subject to ascertainment
bias: facilities enter public documentation largely because residents
complained, so quiet facilities---potentially entire cooling
architectures---are under-represented, and the current record cannot
distinguish ``most facilities are quiet'' from ``most facilities are
unmeasured.'' Fourth, the evidence is asymmetric across operators:
Microsoft, Amazon, and xAI facilities account for most documented
values, while Google, Meta, and Oracle campuses are essentially absent,
so nothing here should be read as ranking companies. Fifth, the
identification of cooling architecture as the dominant design variable
is an engineering inference rather than a measured comparison; the
15--30~dB value is an equipment-level sound-power estimate, not a
measured facility- or residential-receptor difference, and could narrow
substantially under matched-load field measurement. Sixth, the health
pathway is extrapolated from the general low-frequency-noise literature;
no peer-reviewed data-center-specific epidemiological studies were
identified in the sources searched. Finally, the technology is moving
quickly---direct liquid cooling is being adopted at scale for AI
workloads---so the acoustic profile of the newest facilities may differ
from the record summarized here, in either direction.

\section{CONCLUSION}\label{conclusion}

The policy and community-health debate around hyperscale data-center
noise has moved well ahead of its evidence base, and the tools required
to close the gap already exist: the relevant source mechanisms are well
characterized in the HVAC and power-engineering literature, the
measurement and propagation standards are long established, and facility
specifications are increasingly reconstructable from permits and open
hardware documentation. The SYD05 survey demonstrates that a
source-resolved campaign of the required rigor is feasible---conducted,
notably, outside the United States, where the disputes are most acute.
What remains is to carry out equivalent measurement at the U.S.
facilities where communities are already reporting exposure, and to do
so before, rather than after, a health-outcome literature is built on
inadequate exposure data.

\subsection{Availability of Data and
Materials}\label{availability-of-data-and-materials}

Not applicable.

\subsection{Author Contributions}\label{author-contributions}

Davide Bray and Francesca Dominici jointly conceptualized the review,
prepared the original draft, and reviewed and edited the manuscript.

\subsection{Ethics Approval and Consent to
Participate}\label{ethics-approval-and-consent-to-participate}

Not applicable; review article with no human-subject participant data.

\subsection{Acknowledgment}\label{acknowledgment}

None.

\subsection{Financial Support and
Sponsorship}\label{financial-support-and-sponsorship}

None.

\subsection{Conflicts of Interest}\label{conflicts-of-interest}

The authors declare no conflicts of interest.

\section{REFERENCES}\label{references}
\begin{refblock}

\item \refauthors{Johnson K.} Southaven residents raise concerns about noise from xAI turbines {[}Internet{]}. Memphis Flyer; 2025 Nov 20 {[}cited 2026 Aug 13{]}. Available from:
\refurl{https://www.memphisflyer.com/southaven-residents-raise-concerns-about-noise-from-xai-turbines/}

\item \refauthors{Cary P.} Amazon tones down its data center noise after residents sound the alarm {[}Internet{]}. Prince William Times; 2023 Oct 23 {[}cited 2026 Aug 13{]}. Available from:
\refurl{https://www.princewilliamtimes.com/news/amazon-tones-down-its-data-center-noise-after-residents-sound-the-alarm/article_b81154ea-5178-59cb-ac3b-82228a2a52cf.html}

\item \refauthors{Bosker B.} Why everything is getting louder (print edition title: The end of silence) {[}Internet{]}. The Atlantic; 2019 Nov {[}cited 2026 Aug 13{]}. Available from:
\refurl{https://www.theatlantic.com/magazine/archive/2019/11/the-end-of-silence/598366/}

\item \refauthors{Anthony M.} Judson Township residents file lawsuit against Atlas Power, Williams County over data center noise {[}Internet{]}. KFYR-TV; 2023 Nov 29 {[}cited 2026 Aug 13{]}. Available from:
\refurl{https://www.kfyrtv.com/2023/11/29/judson-township-residents-file-lawsuit-against-atlas-power-williams-county-over-data-center-noise/}

\item \refauthors{Pulse White Noise Acoustics.} Microsoft SYD05 data centre: operational noise verification report. Report 250792-R0, prepared for Microsoft Datacentre (Australia) Pty Ltd under New South Wales development consent SSD-10101987 {[}Internet{]}; 2025 Dec 17 {[}cited 2026 Aug 13{]}. Available from:
\refurl{https://local.microsoft.com/wp-content/uploads/2026/05/250792-SYD05-Operational-Noise-Verification-Report-R0.pdf}

\item Musselwhite says xAI complying with city noise level reduction requests {[}Internet{]}. DeSoto Times-Tribune; 2026 May {[}cited 2026 Aug 13{]}. Available from:
\refurl{https://www.desototimes.com/news/musselwhite-says-xai-complying-with-city-noise-level-reduction-requests/article_183a99ed-cbd9-49be-b4c4-1997dd4bfd4a.html}

\item \refauthors{Johnson, Mirmiran \& Thompson (JMT); Nelson Acoustics.} Prince William County noise ordinance update. Presentation to the Prince William County Board of County Supervisors {[}Internet{]}; 2025 Mar 18 {[}cited 2026 Aug 13{]}. Available from:
\refurl{https://www.pwcva.gov/assets/2025-03/DCOAG\%20JMT\%20Noise\%20Ordinance\%20Update\%20Presentation\%2020250318.pdf}

\item Haley et al. v. X.AI Corp. et al. Complaint for damages and demand for jury trial, No. 3:26-cv-00148 (N.D. Miss. filed 2026 Jun 8) {[}Internet{]}. {[}cited 2026 Sep 7{]}. Available from:
\refurl{https://www.mississippifreepress.org/wp-content/uploads/2026/06/Southaven-residents-complaint-against-xAI.pdf}

\item \refauthors{Rozier A, Lin K.} Southaven residents fear pollution, complain of noise, from Elon Musk's xAI data-center turbines {[}Internet{]}. Mississippi Today; 2025 Nov 24 {[}cited 2026 Aug 13{]}. Available from:
\refurl{https://mississippitoday.org/2025/11/24/southaven-residents-fear-pollution-complain-of-noise-from-elon-musks-xai-data-center-turbines/}

\item \refauthors{Redmond M.} Noise remains concern at xAI Southaven location; Action News 5 tests decibel readings {[}Internet{]}. WMC Action News 5; 2026 Jan 7 {[}cited 2026 Aug 13{]}. Available from:
\refurl{https://www.actionnews5.com/2026/01/08/noise-remains-concern-xai-southaven-location-action-news-5-tests-decibel-readings/}

\item \refauthors{Cary P.} Noise levels at Prince William data centers worry opponents of proposed Warrenton facility {[}Internet{]}. Prince William Times; 2022 Aug 31 {[}cited 2026 Aug 13{]}. Available from:
\refurl{https://www.princewilliamtimes.com/noise-levels-at-prince-william-data-centers-worry-opponents-of-proposed-warrenton-facility/article_836f40eb-9a12-52cf-8227-8d5dbf7b1197.html}

\item \refauthors{Browne D.} Great Oak status---DCOAG proposed ordinance levels. Presentation to the Prince William County Data Center Ordinance Advisory Group {[}Internet{]}; 2025 Mar 19 {[}cited 2026 Aug 13{]}. Available from:
\refurl{https://www.pwcva.gov/assets/2025-03/DCOAG\%20Great\%20Oak\%20Status\%20Proposed\%20Ordinance\%20Levels\%20PowerPoint\%2020250319.pdf}

\item \refauthors{Browne D.} AWS data center (Tanner Way) impacts on Great Oak. Presentation to the Prince William County Data Center Ordinance Advisory Group {[}Internet{]}; 2024 Oct 23 {[}cited 2026 Aug 13{]}. Available from:
\refurl{https://www.pwcva.gov/assets/2024-10/Great\%20Oak\%20Update\%20-\%20DCOAG\%2023\%20Oct.pdf}

\item \refauthors{Cary P.} Some cities suffering from data center noise turn to tough limits {[}Internet{]}. Prince William Times; 2023 Feb 27 {[}cited 2026 Aug 13{]}. Available from:
\refurl{https://www.princewilliamtimes.com/news/some-cities-suffering-from-data-center-noise-turn-to-tough-limits/article_2aa7fb8c-b6aa-11ed-8626-5738aae2f76c.html}

\item \refauthors{Dawn-Hiscox T.} CyrusOne forced to take action following noise complaints in Chandler, Arizona {[}Internet{]}. Data Center Dynamics; 2018 Apr 16 {[}cited 2026 Aug 13{]}. Available from:
\refurl{https://www.datacenterdynamics.com/en/news/cyrusone-forced-to-take-action-following-noise-complaints-in-chandler-arizona/}

\item As data center noise concerns grow, Loudoun Co. officials discuss possible mitigation {[}Internet{]}. WUSA9; 2026 Apr 6 {[}cited 2026 Aug 13{]}. Available from:
\refurl{https://www.wusa9.com/article/news/local/data/loudoun-county-officials-data-center-noise/65-45884219-750f-4d47-a660-71a20d8db001}

\item \refauthors{Carey J.} Amid ``constant'' data center noise, Sterling residents also worry about health impact {[}Internet{]}. NBC4 Washington; 2026 Apr 15 {[}cited 2026 Aug 13{]}. Available from:
\refurl{https://www.nbcwashington.com/news/local/northern-virginia/amid-constant-data-center-noise-sterling-residents-also-worry-about-health-impact/4091393/}

\item \refauthors{Holzman J.} I spent the day at a noisy data center. Here's what I learned {[}Internet{]}. Heatmap News (The Fight); 2026 Apr 24 {[}cited 2026 Aug 13{]}. Available from:
\refurl{https://heatmap.news/plus/the-fight/spotlight/data-center-noise}

\item \refauthors{Industrial Noise \& Vibration Centre.} Atlas Power data center environmental noise complaints: \$5 million wasted on predictably ineffective Williston data center noise control {[}Internet{]}. {[}date unknown{]} {[}cited 2026 Aug 13{]}. Available from:
\refurl{https://invc.com/noise-control/atlas-power-data-center-environmental-noise-complaints/}

\item \refauthors{Microsoft Corporation.} Minimizing noise at our datacenter operations {[}Internet{]}. 2022 Oct {[}cited 2026 Aug 13{]}. Available from:
\refurl{https://local.microsoft.com/wp-content/uploads/2022/10/Noise-fact-sheet.pdf}

\item \refauthors{American Society of Heating, Refrigerating and Air-Conditioning Engineers.} 2023 ASHRAE handbook---HVAC applications. Chapter 49: noise and vibration control. Atlanta, GA: ASHRAE; 2023.

\item \refauthors{US Department of Defense.} Unified Facilities Criteria UFC 3-450-01: noise and vibration control {[}Internet{]}. Washington, DC: US Army Corps of Engineers, Naval Facilities Engineering Systems Command, Air Force Civil Engineer Center; 2022 Dec 8 {[}cited 2026 Aug 13{]}. Available from:
\refurl{https://www.wbdg.org/FFC/DOD/UFC/ufc_3_450_01_2022.pdf}

\item \refauthors{Miller L, Dyer I.} Measurements of noise from cooling towers. J Acoust Soc Am 1956;28:758.

\item \refauthors{2CRSi.} What is direct liquid cooling system? {[}Internet{]}. {[}date unknown{]} {[}cited 2026 Aug 13{]}. Available from:
\refurl{https://2crsi.com/direct-liquid-cooling}

\item \refauthors{Kosch L.} Sounds of the data center: quiet of immersion cooling improves quality of life {[}Internet{]}. Data Center Knowledge; 2022 Apr 14 {[}cited 2026 Aug 13{]}. Available from:
\refurl{https://www.datacenterknowledge.com/cooling/sounds-of-the-data-center-quiet-of-immersion-cooling-improves-quality-of-life}

\item \refauthors{Bies DA, Hansen CH.} Engineering noise control: theory and practice. 4th ed. London: Spon Press; 2009.

\item \refauthors{International Electrotechnical Commission.} IEC 60076-10:2016. Power transformers---part 10: determination of sound levels. Geneva: IEC; 2016.

\item \refauthors{International Organization for Standardization.} ISO 1996-1:2016. Acoustics---description, measurement and assessment of environmental noise---part 1: basic quantities and assessment procedures. Geneva: ISO; 2016.

\item \refauthors{Air Movement and Control Association International.} AMCA publication 303-79 (R2012): application of sound power level ratings for fans. Arlington Heights, IL: AMCA International; 2012.

\item \refauthors{Deutsches Institut für Normung.} DIN 45680:1997-03. Messung und Bewertung tieffrequenter Geräuschimmissionen in der Nachbarschaft {[}Measurement and assessment of low-frequency noise immissions in the neighbourhood{]}. Berlin: DIN; 1997. German.

\item \refauthors{International Organization for Standardization.} ISO 1996-2:2017. Acoustics---description, measurement and assessment of environmental noise---part 2: determination of sound pressure levels. Geneva: ISO; 2017.

\item \refauthors{World Health Organization Regional Office for Europe.} Environmental noise guidelines for the European region. Copenhagen: WHO Regional Office for Europe; 2018.

\item \refauthors{International Organization for Standardization.} ISO 9613-2:2024. Acoustics---attenuation of sound during propagation outdoors---part 2: engineering method for the prediction of sound pressure levels outdoors. 2nd ed. Geneva: ISO; 2024.

\item \refauthors{Moorhouse AT, Waddington DC, Adams MD.} Procedure for the assessment of low frequency noise complaints. Defra contract NANR45, revision 1. Salford: University of Salford, Acoustics Research Centre; 2011.

\item \refauthors{Leventhall G, Pelmear P, Benton S.} A review of published research on low frequency noise and its effects. Report for the Department for Environment, Food and Rural Affairs. London: Defra; 2003.

\item \refauthors{Guski R, Schreckenberg D, Schuemer R.} WHO environmental noise guidelines for the European region: a systematic review on environmental noise and annoyance. Int J Environ Res Public Health 2017;14:1539.

\item \refauthors{Basner M, McGuire S.} WHO environmental noise guidelines for the European region: a systematic review on environmental noise and effects on sleep. Int J Environ Res Public Health 2018;15:519.

\item \refauthors{van Kempen E, Casas M, Pershagen G, Foraster M.} WHO environmental noise guidelines for the European region: a systematic review on environmental noise and cardiovascular and metabolic effects: a summary. Int J Environ Res Public Health 2018;15:379.

\item \refauthors{Baliatsas C, van Kamp I, van Poll R, Yzermans J.} Health effects from low-frequency noise and infrasound in the general population: is it time to listen? A systematic review of observational studies. Sci Total Environ 2016;557-558:163-9.

\item \refauthors{Leventhall HG.} Low frequency noise and annoyance. Noise Health 2004;6:59-72.

\item \refauthors{Persson Waye K.} Effects of low frequency noise on sleep. Noise Health 2004;6:87-91.
\end{refblock}

\end{document}